\documentclass[prb,aps,eqsecnum,twocolumn,amssymb,showpacs]{revtex4}
\usepackage{color}
\usepackage{graphicx}
\usepackage{longtable}
\usepackage{epsfig}
\usepackage{dcolumn}
\usepackage{bm}

\usepackage{amssymb}

\begin{document}

\title{        Exact solution for a quantum compass ladder }

\author{     Wojciech Brzezicki}
\affiliation{Marian Smoluchowski Institute of Physics, Jagellonian
             University, Reymonta 4, PL-30059 Krak\'ow, Poland }

\author {    Andrzej M. Ole\'{s} }
\affiliation{Marian Smoluchowski Institute of Physics, Jagellonian
             University, Reymonta 4, PL-30059 Krak\'ow, Poland \\
             Max-Planck-Institut f\"ur Festk\"orperforschung,
             Heisenbergstrasse 1, D-70569 Stuttgart, Germany  }

\date{\today}

\begin{abstract}
We introduce a spin ladder with antiferromagnetic Ising ZZ
interactions along the legs, and interactions on the rungs which
interpolate between the Ising ladder and the quantum compass
ladder. We show that the entire energy spectrum of the ladder may
be determined exactly for finite number of spins $2N$ by mapping
to the quantum Ising chain and using Jordan-Wigner transformation
in invariant subspaces. We also demonstrate that subspaces with
spin defects lead to excited states using finite size scaling, and
the ground state corresponds to the quantum Ising model without
defects. At the quantum phase transition to maximally frustrated
interactions of the compass ladder, the ZZ spin correlation
function on the rungs collapses to zero and the ground state
degeneracy increases by 2. We formulate a systematic method to
calculate the partition function for a mesoscopic system, and
employ it to demonstrate that fragmentation of the compass ladder
by kink defects increases with increasing temperature. The
obtained heat capacity of a large compass ladder consisting of
$2N=104$ spins reveals two relevant energy scales and has a broad
maximum due to dense energy spectrum. The present exact results
elucidate the nature of the quantum phase transition from ordered
to disordered ground state found in the compass model in two
dimensions.

{\it Published in: Phys. Rev. B {\bf 80}, 014405 (2009).}
\end{abstract}


\maketitle

\section{Introduction}
\label{sec:intro}

Spin ladders play an important role in quantum magnetism. Interest
in them is motivated by their numerous experimental realizations
in transition metal oxides \cite{Dag99} and has increased over the
last two decades. One of recently investigated realizations of
spin ladders are Sr$_{n-1}$Cu$_{n+1}$O$_{2n}$ cuprates (with
$n=3,5,7,\cdots$),\cite{Gop94} and the simplest of them, a spin
ladder with two legs connected by rungs, is realized in
Sr$_2$Cu$_4$O$_6$. Excitation spectra of such antiferromagnetic
(AF) spin ladders are rich and were understood only in the last
decade. They consist of triplet excitations, bound states and
two-particle continuum,\cite{Tre00} and were calculated in
unprecedented detail for quantum AF spin $S=1/2$ two-leg ladder
employing optimally chosen unitary transformation.\cite{Kne01} In
some of spin ladder systems charge degrees of freedom also play a
role, as for instance in $\alpha'$-NaV$_2$O$_5$, where AF order
and charge order coexist in spin ladders with two
legs,\cite{Hor98} or in the Cu--O planes of
La$_x$Sr$_{14-x}$Cu$_{24}$O$_{41}$, where spin and charge order
coexist for some values of $x$.\cite{Woh07} This advance in the
theoretical understanding of the ground states and excitation
spectra of spin ladders is accompanied by recent experimental
investigations of triplon spectra by inelastic neutron scattering
\cite{Not07} of almost perfect spin ladders in
La$_4$Sr$_{10}$Cu$_{24}$O$_{41}$. Finally, in the theory spin
ladders could serve as a testing ground for new (ordered or
disordered) phases which might arise for various frustrated
exchange interactions.\cite{Pen07}

A particularly interesting situation arises when frustration of
spin interactions may be tuned by varying strength of certain
coupling constants, and could thus exhibit transitions between
ordered and disordered phases. On the one hand, periodically
distributed frustrated Ising interactions do not suffice to
destroy magnetic long-range order in a two-dimensional (2D)
system, but only reduce the temperature of the magnetic phase
transition.\cite{Lon80} On the other hand, when the model is
quantum, increasing frustration of exchange interactions may
trigger a quantum phase transition (QPT), as for instance in the
one-dimensional (1D) compass model.\cite{Brz07} Physical
realizations of frustrated interactions occur in 2D and
three-dimensional spin-orbital models derived for Mott insulators
in transition metal oxides in the orbital part of the
superexchange. In such models frustration is intrinsic and follows
from the directional nature of orbital interactions.\cite{Fei97}
Usually such frustration is removed either by Hund's exchange
$J_H$ or by Jahn-Teller orbital interactions, but when these terms
are absent it leads to a disordered orbital liquid ground state.
Perhaps the simplest realistic example of this behavior is the
(Kugel-Khomskii) model for Cu$^{2+}$ ions in $d^9$ electronic
configuration at $J_H=0$, where a disordered ground state was
found.\cite{Fei98} Examples of such disordered states are either
various valence-bond phases with singlet spin configurations on
selected bonds,\cite{Mat04} or orbital liquids established both in
$t_{2g}$ systems\cite{Kha00} and in $e_g$ systems.\cite{Fei05}
Characteristic features of spin-orbital models are enhanced
quantum effects and entanglement, \cite{Ole06} so their ground
states cannot be predicted using mean-field decoupling schemes.
Also in doped systems some unexpected features emerge for
frustrated orbital superexchange interactions, and the
quasiparticle states are qualitatively different from those
arising in the spin $t$--$J$ model. \cite{vdB00} Therefore, it is
of great interest to investigate spin models with frustrated
interactions which stand for the orbital part of the
superexchange, particularly when such models could be solved
exactly.

Although the orbital superexchange interactions are frequently
Ising-like, they lead to quantum models with intrinsically
frustrated exchange models as different orbital components
interact depending on the bond orientation in real
space.\cite{vdB04} A generic case of such frustrated interactions
is the so-called 2D quantum compass model, \cite{Kho03} which was
recently investigated numerically. \cite{Mil05,Wen08} Although
orbital superexchange interactions in Mott insulators are
typically AF,\cite{Fei97,Fei98,Mat04,Kha00} a similar frustration
concerns also ferromagnetic (FM) interactions, and a QPT was also
found in the compass model with FM interactions.\cite{Che07}

The 1D variant of the compass model with alternating interactions
of $z$-th and $x$-th spin components on even and odd bonds was
solved exactly by an analytical method,\cite{Brz07} and
entanglement in the ground state was analyzed
recently.\cite{You08} We note that the 1D compass model (the model
of Ref. \onlinecite{Brz07} in the limit of equal and alternating
interactions on the bonds) is equivalent to the 1D anisotropic XY
model, solved in the seventies.\cite{Per75} An exact solution of
the 1D compass model demonstrates that certain nearest-neighbor
spin correlation functions change discontinuously at the point of
a QPT when both types of interactions have the same strength. This
somewhat exotic behavior follows because the QPT occurs at the
multicritical point in the parameter space.\cite{Eri09} A similar
discontinuous behavior of nearest-neighbor spin correlations was
also found numerically for the 2D compass model.\cite{Kho03} While
small anisotropy of interactions leads to particular short-range
correlations dictated by the stronger interaction, in both 1D and
2D compass model one finds a QPT to a highly degenerate disordered
ground state when the interactions are balanced.

The purpose of this paper is to present an exact solution of the
compass model on a spin ladder, with ZZ Ising interactions between
$z$-th spin components along the ladder legs, and interactions on
the rungs which gradually evolve from ZZ Ising interactions to XX
Ising ones. In this way the interactions interpolate between the
classical Ising spin ladder and the quantum compass ladder with
frustrated interactions. The latter case will be called compass
ladder below --- it stands for a generic competition between
orbital interactions on different bonds and can serve to
understand better the physical consequences of the frustrated
orbital superexchange.

The paper is organized as follows. The model and its invariant
dimer subspaces are introduced in Sec. II. Next the ground state
and the lowest excited states of the model are found in Sec. III
by solving the model in all nonequivalent subspaces. Thereby we
discuss the role played by defects in spin configuration and show
that the ground state is obtained by solving the 1D quantum Ising
(pseudospin) model (QIM). Using an example of a finite system, we
provide an example of the energy spectrum, and next extrapolate
the ground state energy obtained for finite systems to the
thermodynamic limit. We also present the changes of spin
correlations at the QPT, and derive the long-range spin
correlations. Next we construct canonical ensemble for the spin
ladder in Sec. IV and present the details concerning the
calculation of energies in the appendix. The constructed partition
function is used to derive such thermodynamic properties of the
compass ladder as the temperature variation of spin correlations,
and the average length of fragmented chains separated by kinked
areas in Sec. V. In Sec. \ref{sec:cv} we present the evolution of
heat capacity $C_V$ when interactions change from the Ising to
compass ladder for a small ladder of $N=8$ spins, and next analyze
$C_V$ for a large (mesoscopic) compass ladder of $2N=104$ spins.
While the characteristic excitation energies responsible for the
maxima in heat capacities can be deduced from the energy spectrum
for $N=8$ spins, generic features of excitations follow from the
form of $C_V$ in case of the mesoscopic compass ladder. Final
discussion and summary of the results are given in Sec. VII.

\section{Compass model on a ladder}
\label{sec:cola}

\begin{figure}[t!]
\includegraphics[width=7.2cm,angle=0]{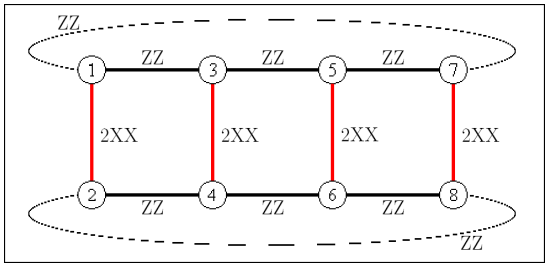}
\caption{(Color online) Schematic view of the quantum compass
ladder with $N=4$ rungs, described by Hamiltonian (\ref{cola})
with $\alpha=1$. Interactions along the ladder legs labeled as
$ZZ$ (horizontal lines) are $\sigma^z_{2i-1}\sigma^z_{2i+1}$
(upper leg) and $\sigma^z_{2i}\sigma^z_{2i+2}$ (lower leg). The
interactions along the rungs labeled as $2XX$ (vertical lines) are
$2\sigma^x_{2i-1}\sigma^x_{2i}$ (the factor of 2 simulates the
periodic boundary condition along the rungs). Dashed lines
indicate periodic boundary conditions along the ladder legs. }
\label{fig:colade}
\end{figure}

We consider a spin ladder with $N$ rungs $\langle 2i-1,2i\rangle$
labelled by $i=1,2,\cdots,N$. The interactions along ladder legs
are Ising-like with AF coupling $J$ between $z$-th spin components
($\sigma_i^z\sigma_{i+1}^z$), while AF interactions on the rungs
interpolate between the Ising coupling of $z$-th
($2\sigma_{n-1}^z\sigma_{n+1}^z$) and $x$-th
($2\sigma_{n-1}^x\sigma_{n+1}^x$) spin components,
\begin{eqnarray}
{\cal H}(\alpha)&=&2J \sum_{i=1}^{N}\left\{
 \alpha \sigma^x_{2i-1}\sigma^x_{2i}
+(1-\alpha)\sigma^z_{2i-1}\sigma^z_{2i} \right\} \nonumber \\
&+&J\sum_{i=1}^{N}
\left(\sigma^z_{2i-1}\sigma^z_{2i+1}+\sigma^z_{2i}\sigma^z_{2i+2}
\right)\,,
\label{cola}
\end{eqnarray}
by varying parameter $0\leq\alpha\leq 1$. We assume periodic
boundary conditions along the ladder legs, i.e.,
$\sigma^z_{2N+1}\equiv\sigma^z_1$ and
$\sigma^z_{2N+1}\equiv\sigma^z_2$. The factor of two for the
interactions on the rungs $\propto 2J$ was chosen to guarantee the
same strength of interactions on the rungs (with only one rung
neighbor of each spin) as along the ladder legs (with two leg
neighbors). Increasing $\alpha$ gradually modifies the
interactions on the rungs and increases frustration. For
$\alpha=0$ one finds the reference Ising ladder, while at
$\alpha=1$ the interactions describe a competition between
frustrated ZZ interactions along the ladder legs and 2XX
interactions on the rungs, characteristic of the compass ladder. A
representative compass ladder with $N=4$ rungs (i.e., $2N=8$
spins) is shown in Fig. \ref{fig:colade}.

To solve the spin ladder given by Eq. (\ref{cola}) in the range of
$0\leq\alpha\leq 1$ we notice that $[{\cal
H}(\alpha),\sigma_{2i-1}^z\sigma_{2i}^z]\equiv 0$. Therefore we
have a set of $N$ symmetry operators,
\begin{equation}
R_i\equiv\sigma_{2i-1}^z\sigma_{2i}^z\,,
\label{ri}
\end{equation}
with respective eigenvalues $r_i=\pm 1$. Each state of the system
can be thus written in a basis of $\sigma^z_i$ eigenvectors
$\left|s_{1},s_{2},s_{3}, \ldots,s_{2N}\right\rangle$ fixed by
strings of quantum numbers $s_i=\pm 1$. These vectors can be
parametrized differently by a new set of quantum numbers $\{t_i\}$
and $\{r_i\}$, with $i=1,2,\cdots,N$; they are related to the old
ones by the formulae: $t_{i}\equiv s_{2i-1}$ and $r_{i}\equiv
s_{2i-1}s_{2i}$. Now we introduce new notation for the basis
states
\begin{equation}
\left|t_{1},t_{2},\ldots,t_{N}\right\rangle _{r_{1}r_{2}\cdots r_{N}}
\equiv\left|t_{1},t_{1}r_{1},t_{2},t_{2}r_{2},\ldots ,t_{N},t_{N}r_{N}
\right\rangle\ , \label{sub}
\end{equation}
where the right-hand side of Eq. (\ref{sub}) is the state
$\left|s_{1},s_{2},s_{3},\ldots,s_{2N}\right\rangle$ written in
terms of variables $\{t_i\}$ and $\{r_i\}$, and the left-hand side
defines new notation. This notation highlights the different role
played by $r_i$'s, which are conserved quantities, and by $t_i$'s,
being new pseudospin variables. For states like in Eq.
(\ref{sub}), we define new pseudospin operators $\tau_{i}^{z}$ and
$\tau_{i}^x$ acting on $\{t_i\}$ quantum numbers as Pauli
matrices, e.g. for $i=1$:
\begin{eqnarray}
\tau_{1}^x |t_{1},t_{2},\ldots,t_{N}\rangle _{r_{1}r_{2}\cdots
r_{N}}& = & |-t_{1},t_{2},\ldots,t_{N}\rangle _{r_{1}r_{2}\cdots
r_{N}}\ ,
\nonumber \\
\tau_{1}^{z} |t_{1},t_{2},\ldots,t_{N}\rangle _{r_{1}r_{2}\cdots
r_{N}}& = & t_{1} |t_{1},t_{2},\ldots,t_{N}\rangle
_{r_{1}r_{2}\cdots r_{N}}\ . \nonumber \\
\end{eqnarray}
A similar transformation was introduced for a frustrated
spin-$1/2$ chain by Emery and Noguera,\cite{Eme88} who showed that
it can by mapped onto an Ising model in a transverse field.
Recently this procedure was used to investigate quantum
criticality in a two-leg strongly correlated ladder model at
quarter filling.\cite{Laa08}

The Hamiltonian can be now written in a common eigenbasis of $R_i$
(\ref{ri}) operators by means of $\{\tau_i^x,\tau_i^z\}$ operators.
In a subspace labelled by a string $r_1,r_2,\cdots,r_N$, the
reduced form of the Hamiltonian is
\begin{eqnarray}
{\cal H}_{r_{1}r_{2}\cdots r_{N}}(\alpha)&\equiv &
J\sum_{i=1}^{N}\left\{(1+r_ir_{i+1})\tau_{i}^z\tau_{i+1}^z
+2\alpha\tau_{i}^x\right\} \nonumber \\
&+& 2JC_{\vec r}(\alpha)\ , \label{hamef}
\end{eqnarray}
with a constant
\begin{equation}
C_{\vec r}(\alpha)=(1-\alpha)\sum_{i=1}^Nr_i\,, \label{cr}
\end{equation}
and periodic boundary condition $\tau_{N+1}^z\equiv\tau_{1}^z$.
This leads to the exactly solvable QIM with transverse
field,\cite{Lie61,Dzi05,Mat06} if only $r_i\equiv 1$ or $r_i\equiv
-1$. Otherwise there are always some $\tau^z_i\tau^z_{i+1}$
interactions missing (defects created in the chain) and we obtain
a set of disconnected quantum Ising chains with loose ends and
different lengths. The bonds with no pseudospin interactions may
stand next to each other, so in an extreme case when
$r_{i+1}=-r_i$ for all $i$, one finds no Ising bonds and no chains
appear.


One may easily recognize that the ground state of the spin ladder
described by Hamiltonian (\ref{cola}) lies in a subspace with
$r_i\equiv -1$ for $\alpha<1$. First of all, $r_i\equiv -1$
minimizes $C_{\vec r}(\alpha)$, see Eq. (\ref{cr}). To understand
a second reason which justifies the above statement let us examine
a partial Hamiltonian (open chain) of the form
\begin{equation}
{\cal H}(\alpha,L)=2J \sum_{i=1}^{L-1}\tau_{i}^z\tau_{i+1}^z
+2J\alpha\sum_{i=1}^L\tau_{i}^x\ ,
\label{chain}
\end{equation}
with $2\leq L\leq N-1$. Note that it appears generically in Eq.
(\ref{hamef}) and consists of two terms containing pseudospin
operators $\{\tau^x_i\}$ and $\{\tau^z_i\}$. Let us call them
${\cal H}^x$ and ${\cal H}^z$ and denote the ground state of
${\cal H}^x$ as $|x\rangle$ with energy $E_x$. The mean value of
${\cal H}(\alpha,L)$ in state $|x\rangle$ is also $E_x$ because
every $\tau_i^z$ operator has zero expectation value in state
$|x\rangle$, i.e., $\langle x|\tau_i^z|x\rangle=0$. However, we
know that $|x\rangle$ is not an eigenvector of ${\cal
H}(\alpha,L)$ which implies that ${\cal H}(\alpha,L)$ must have a
lower energy than $E_x$ in the ground state. This shows that the
presence of $\tau_{i}^z\tau_{i+1}^z$ bonds in the Hamiltonian
${\cal H}(\alpha,L)$ lowers the energy of bare ${\cal H}^x$. One
may also expect that this energy decreases with increasing length
$L$ of the chain, and is proportional to $L$ in the thermodynamic
limit. The numerical evidence for this are plots of the ground
state energy versus $L$ presented in section 3. Looking at
Hamiltonian (\ref{hamef}) we see that the longest chains of the
type (\ref{chain}) appear in subspaces with $r_i\equiv -1$ and
$r_i\equiv 1$, but the constant term $C_{\vec r}(\alpha)$ favors
$r_i\equiv -1$ if only $\alpha<1$. For $\alpha=1$ the ground state
can be in both subspaces, and its degeneracy follows, see below.

\section{Energy spectra in invariant subspaces}
\label{sec:ex}

\subsection{Quantum Ising model}
\label{sec:qim}

To find the ground state of spin ladder (\ref{cola}) we need to
solve the QIM that arises from Eq. (\ref{hamef}) when $r_i\equiv
-1$. Thus we need to diagonalize the Hamiltonian of the form
\begin{equation}
{\cal H}_{\rm QIM}(\beta,\alpha)=2J\sum_{i=1}^{N}
(\beta\tau_{i}^z\tau_{i+1}^z+\alpha\tau_{i}^x)\ ,
\label{QIM}
\end{equation}
which is related to our original problem by the formula
\begin{equation}
{\cal H}_{-1-1\cdots -1}={\cal H}_{\rm QIM}(1,\alpha)-2NJ(1-\alpha)\ .
\label{org}
\end{equation}
The formal parameter $\beta$ is introduced for convenience and
will be used to determine the correlation functions along the
ladder legs by differentiation, see below. The standard way of
solving ${\cal H}_{\rm QIM}$ starts with Jordan--Wigner (JW)
transformation. This non--linear mapping replacing spin operators
by spinless fermions is of the form
\begin{eqnarray}
\tau_{j}^z&=&(c_{j}^{}+c_{j}^{\dagger})
{\prod_{i<j}}(1-2c_{i}^{\dagger}c_{i})\ , \nonumber \\
\tau_{j}^x&=&(1-2c_{j}^{\dagger}c_{j}^{})\ . \label{JWT}
\end{eqnarray}
The boundary condition for fermion operators $\{c_i\}$ after
inserting them into ${\cal H}_{\rm QIM}$ (\ref{QIM}) is
antiperiodic for even and periodic for odd number of JW
quasiparticles in the chain. The operator ${\cal P}$ of the parity
of fermions,
\begin{equation}
{\cal P}=\prod_{i=1}^N\,(1-2c_{i}^{\dagger}c_{i})\ , \label{pari}
\end{equation}
corresponds to the operation of flipping all spins along the
$z$-th axis and commutes with ${\cal H}_{\rm QIM}$. Therefore, the
Hamiltonian can be split into two diagonal blocks, for even $(+)$
and odd $(-)$ number of JW fermions by means of projection
operators $\frac{1}{2}(1\pm {\cal P})$. Therefore we write
\begin{equation}
{\cal H}_{\rm QIM}=\frac{1}{2}(1+{\cal P}){\cal H}^+
+\frac{1}{2}(1-{\cal P}){\cal H}^-\ , \label{Proj}
\end{equation}
where
\begin{eqnarray}
{\cal H}^{\pm}&=&2J\sum_{i=1}^N\left\{\beta(c_{i}^{\dagger}-c_{i})
(c_{i+1}^{\dagger}+c_{i+1})-2\alpha
c_{i}^{\dagger}c_{i}\right\} \nonumber \\
&+&2JN\alpha\ , \label{qimJW}
\end{eqnarray}
with two different boundary conditions: $c_{N+1}=\mp c_1$ for
($\pm$) subspaces. Let us point out that the only consequence of
the nonlinearity of the JW transformation is the minus sign which
appears in the first bracket multiplying $\beta$. This is thanks
to one--dimensionality and only nearest-neighbor interactions in
the reduced Hamiltonian (\ref{hamef}), but is not the case for the
original Hamiltonian (\ref{cola}).

Next step is the Fourier transformation,
\begin{equation}
c_j=\frac{1}{\sqrt{N}}\sum_{k}e^{ijk}c_k\,, \label{fouc}
\end{equation}
with quasimomenta $k=\pm(2l-1)\pi/N$ $[l=1,2,\cdots,N/2]$ in an
even subspace $(+)$, and $k=0,\pi,\pm 2l\pi/N$
$[l=1,2,\cdots,(N/2-1)]$ in an odd one $(-)$. After transforming
the operators in Eq. (\ref{qimJW}) we obtain ${\cal H}^{\pm}$ in a
block diagonal form,
\begin{eqnarray}
{\cal H}^{\pm}&=&4J{\sum _k}^{\pm}(\beta\cos k
-\alpha)c^\dagger_k c^{}_k \nonumber \\
&+&2J{\sum _k}^{\pm}\beta(c^\dagger_kc^\dagger
_{-k}e^{ik}+h.c.)+2JN\alpha\,. \label{Hfouc}
\end{eqnarray}
Diagonalization is completed by a Bogoliubov transformation,
defining new fermion operators ${\gamma}^\dagger_k\equiv\alpha_k
c^\dagger_k+\beta_k c_{-k}$ (for $k\neq 0,\pi$, while the
operators $c_0$ and $c_\pi$ have no partner and are left
untransformed). Transformation coefficients $\alpha_k$ and
$\beta_k$ are obtained from the condition
\begin{equation}
\left[{\cal H}_{\rm
QIM},\gamma^\dagger_k\right]=\omega_k\gamma^\dagger_k\,,
\label{comm}
\end{equation}
which is an eigenproblem in linear space spanned by operators
$c_k^{\dagger}$ and $c_{-k}$. We get two eigenvectors
$(\alpha_k,\beta_k)$, corresponding to the quasiparticle operators
${\gamma}^\dagger_k$ and ${\gamma}_{-k}$, and two corresponding
eigenvalues $\omega_k=\pm E_k$, with
\begin{equation}
E_k(\beta,\alpha)=4J\left\{\alpha^2+\beta^2 - 2\alpha\beta\cos
k\right\}^{1/2}\,. \label{Ek}
\end{equation}
Therefore, the Hamiltonian is brought to the diagonal form in both
subspaces
\begin{eqnarray}
\label{qimD}
{\cal H}^{+}&=&{\sum_k}^+
E_k\left(\gamma^\dagger_{k}\gamma^{}_{k}-\frac{1}{2}\right)\ ,  \\
{\cal H}^{-}&=&{\sum_k}^-
E_k\left(\gamma^\dagger_{k}\gamma^{}_{k}-\frac{1}{2}\right)
+4J(\beta-\alpha)c^{\dagger}_0 c_0 \nonumber \\
&-&4J(\beta+\alpha)c^\dagger_\pi
c^{}_\pi+4J\alpha\ .
\end{eqnarray}

We still need to transform the parity operator ${\cal P}$.
Luckily, the Fourier transformation does not change its form and
to see that so does the Bogoliubov transformation, one can look at
the vacuum state $|0\rangle$ for quasiparticle operators
$\gamma_k$. From the condition $\gamma_k |0\rangle=0$ for all $k$
we get
\begin{equation}
|0\rangle=\prod_k\,\left(\bar {\alpha}_k+\bar{\beta}_k
c_{-k}^{\dagger}c_{k}^{\dagger}\right)|vac\rangle\ ,
\end{equation}
where $|vac\rangle$ is a true vacuum state for JW fermions or a
state with all spins up. From the form of $|0\rangle$ we see that
it contains a superposition of all even numbers of quasiparticles
$c_k^{\dagger}$, and the total quasiparticle number is not fixed.
Acting on the vacuum with a single creation operator
$\gamma_k^{\dagger}$ we obtain a state with odd number of JW
fermions, because $\gamma_k^{\dagger}$ is a linear combination of
a creation $c_k^{\dagger}$ and annihilation $c_{-k}$ operator of a
single fermion. In this way one may get convinced that the parity
of quasiparticles $\gamma_k^{\dagger}$ and $c_k^{\dagger}$ is the
same.

\subsection{Ground state and the energy spectrum }
\label{sec:ene}

\begin{figure}[t!]
\includegraphics[width=8cm,angle=0]{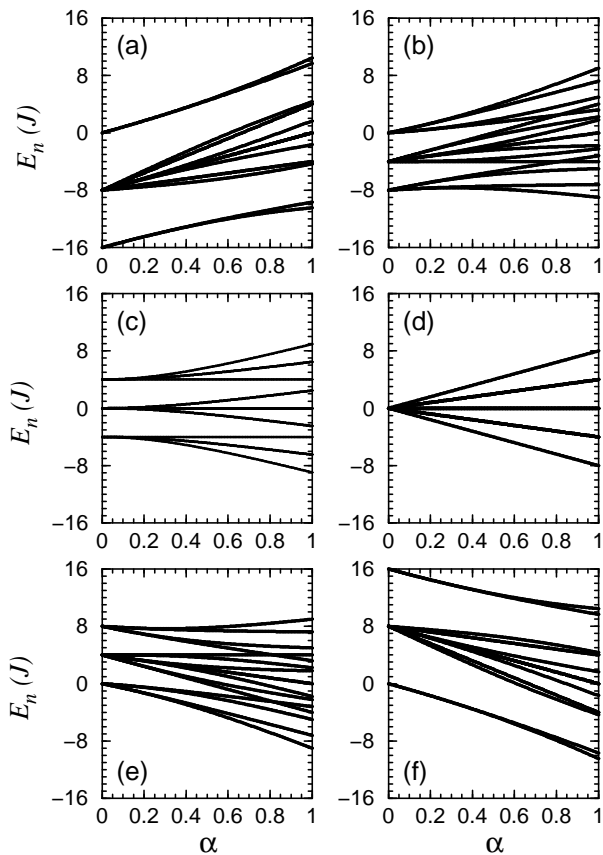}
\caption{ Eigenenergies $E_n$ of the spin ladder (\ref{cola}) of
Fig. \ref{fig:colade} with $N=4$ rungs for increasing $\alpha$,
obtained by exact diagonalization. Different panels show energies
in invariant subspaces of the effective Hamiltonian (\ref{hamef}),
with $1$ and $\bar 1$ standing for positive or negative values of
$r_i$: (a) $\bar 1\bar 1\bar 1\bar 1$, (b) $1\bar 1\bar 1\bar 1$,
(c) $11\bar 1\bar 1$, (d) $1 \bar 11 \bar 1$, (e) $111\bar 1$ and
(f) $1111$. While the subspaces (a) and (f) are unique, other
subspaces are equivalent by symmetry to those shown in panels
(b)--(e), resulting in total spectrum of 256 eigenstates. Quantum
phase transition occurs at $\alpha=1$, where the lowest
eigenenergies in the subspaces (a) and (f) become degenerate. In
the thermodynamic limit $N\to\infty$ the spectrum changes
qualitatively --- the two lowest energies in the subspaces $\bar
1\bar 1\bar 1\bar 1$ and $1111$ are degenerate and the ground
state from the subspace (b) ($1\bar 1\bar 1\bar 1$) becomes the
first excited state of the spin ladder. } \label{fig:ene}
\end{figure}

From the diagonal form of the QIM Hamiltonian given by Eq.
(\ref{qimD}) we see that the ground state of spin ladder
(\ref{cola}) is simply $|0\rangle$ in subspace $r_i\equiv -1$ (or
$r_i\equiv 1$ when $\alpha=0$). For the ground state energy, one
uses Eq. (\ref{org}) to get
\begin{equation}
E_{-1-1\cdots -1}=E_{\rm QIM}(1,\alpha)-2NJ(1-\alpha)\ ,
\end{equation}
with $E_{\rm QIM}(1,\alpha)$ given in the thermodynamic limit by
an integral
\begin{equation}
E_{\rm QIM}(\beta,\alpha)=-\frac{N}{2\pi}\int_{0}^{\pi}
dk\;E_k(\beta,\alpha) \,. \label{eqim}
\end{equation}
The ground state in the absence of transverse field (at
$\alpha=0$) is doubly degenerate --- it is given by two possible
N\'eel states. At finite $\alpha>0$, this degeneracy is removed,
and the sum of the two N\'eel states (symmetric state),
$|0_+\rangle$, is the ground state, while their difference
(antisymmetric state) becomes the first excited state. This first
excited state, $|0_-\rangle=\gamma_{\pi}^{\dagger}|0_+\rangle$,
stems from the same subspace and belongs to the spectrum of ${\cal
H}^-$. The splitting of the states $|0_+\rangle$ and $|0_-\rangle$
increases with $\alpha$, see Fig. \ref{fig:ene}(a). For finite $N$
and $\alpha>0$ there is always finite energy difference between
the energies of $|0_+\rangle$ and
$|0_-\rangle=\gamma_{\pi}^{\dagger}|0_+\rangle$ states. However,
in the thermodynamic limit $N\to\infty$, this energy gap vanishes
for $\alpha\leq 1$ and starts to grow as $4J\alpha$ at $\alpha=1$.

The full spectrum for the ladder with $N=4$ rungs belongs to six
classes of subspaces equivalent by symmetry --- it is depicted in
Fig. \ref{fig:ene}. With increasing $\alpha$ the spectrum changes
qualitatively from discrete energy levels of the classical Ising
ladder at $\alpha=0$, with the ground state energy per spin equal
$-2J$, to a narrower and quasi--continuous spectrum when the
quantum compass ladder at $\alpha=1$ is approached, with the
ground state energy $-4J/\pi$ per spin. At the $\alpha=1$ point
one finds an additional symmetry; subspaces indexed by $\vec r$
and $-\vec r$ are then equivalent which makes each energy level at
least doubly degenerate.

\subsection{Correlation functions}
\label{sec:corr}

\begin{figure}[t!]
\includegraphics[width=7cm,angle=0]{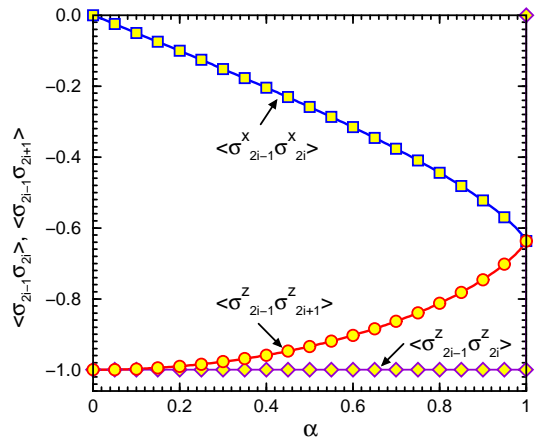}
\caption{(Color online) Nearest neighbor correlation functions in
the ground state for spin ladder (\ref{cola}) in the thermodynamic
limit $N\to\infty$. For increasing $\alpha$ spin correlations
$\langle \sigma_{2i-1}^x\sigma_{2i}^x\rangle$ on the rungs
decrease from zero to $-2/\pi$. At the same time, AF correlations
$\langle\sigma_{2i-1}^z\sigma_{2i+1}^z\rangle$ along the ladder
legs gradually weaken (increase from the classical value $-1$ at
$\alpha=0$ to $-2/\pi$ at $\alpha=1$), and become degenerate with
the rung $\langle\sigma_{2i-1}^x\sigma_{2i}^x\rangle$ correlations
at the quantum critical point $\alpha=1$. Correlation function
$\langle\sigma_{2i-1}^z\sigma_{2i}^z\rangle$ on the rungs,
directly related to the subspace indices $r_i$, remains constant
$(\langle\sigma_{2i-1}^z\sigma_{2i}^z\rangle=-1)$ in the entire
range of $\alpha<1$, and jumps to $0$ at $\alpha=1$. }
\label{fig:corr}
\end{figure}

All the nontrivial nearest neighbor spin correlation functions in
the ground state can be determined by taking derivatives of the
ground state energy $E_{\rm QIM}(\beta,\alpha)$ (\ref{eqim}) with
respect to $\alpha$ or $\beta$, while the others are evident from
the construction of the subspaces. In this way one finds
$\langle\sigma^z_{2i-1}\sigma^z_{2i+1}\rangle$ correlation along
the legs and $\langle\sigma^x_{2i-1}\sigma^x_{2i}\rangle$ along
the rungs, shown in Fig. \ref{fig:corr}. Spin correlations
$\langle\sigma^z_{2i-1}\sigma^z_{2i+1}\rangle$ along the legs
increase from the classical value $-1$ up to $-2/{\pi}$ for
$\alpha=1$. By symmetry, both ladder legs are equivalent and
$\langle\sigma^{\alpha}_{2i-1}\sigma^{\alpha}_{2i+1}\rangle=
\langle\sigma^{\alpha}_{2i}\sigma^{\alpha}_{2i+2}\rangle$ for
$\alpha=x,z$. At the same time spin correlations
$\langle\sigma^x_{2i-1}\sigma^x_{2i}\rangle $ along the rungs
gradually develop from $0$ in the classical limit to $-2/{\pi}$ at
the quantum critical point $\alpha=1$. Both functions meet at
$\alpha=1$ which indicates balanced interactions --- ZZ along the
legs and 2XX along the rungs in case of the quantum compass ladder
(see Fig. \ref{fig:colade}).

For the remaining correlations one finds
\begin{eqnarray}
\label{ssxi}
\langle\sigma^x_{2i-1}\sigma^x_{2i+1}\rangle &=&0\ , \\
\label{sszi} \langle\sigma^z_{2i-1}\sigma^z_{2i}\rangle&=&\langle
R_i\rangle=r_i\ .
\end{eqnarray}
Eq. (\ref{ssxi}) follows from the fact that operators
$\sigma^x_{2i-1}\sigma^x_{2i+1}$ do not commute with the symmetry
operators $R_i$ (\ref{ri}). In turn, averages of the symmetry
operators along the rungs (\ref{sszi}) are constant and equal $-1$
for $\alpha<1$, but at $\alpha=1$ they change in a discontinuous
way and become zero, because at this point the degeneracy of the
ground state increases to $2\times 2=4$, and the spins on the
rungs are disordered, so the ZZ correlations vanish.

Finally, one can calculate the long range correlation functions
for $z$-th spin components,
\begin{equation}
\label{sszl} \langle\sigma^z_{2i+a}\sigma^z_{2j+b}\rangle=
r_i^{a+1}r_j^{b+1}\langle\tau^z_i\tau^z_j\rangle\ .
\end{equation}
The right--hand side of Eq. (\ref{sszl}) can be obtained from the
QIM by the so--called Toeplitz determinant \cite{Mat06} and can be
also found in Ref. \onlinecite{Brz07}. All the long range XX
correlation functions are zero in the ground state as they do not
commute with $R_i$'s operators (\ref{ri}).

Note that correlations $\langle\tau^z_i\tau^z_j\rangle $ vanish in
any subspace when $|i-j|$ exceeds the length of the longest Ising
chain. This is due to the fact that, as already mentioned in
section \ref{sec:cola}, the effective Hamiltonian in a given
subspace describes a set of completely independent quantum Ising
chains. Thus, at finite temperature, one can expect that the
compass ladder will be {\it more} disordered than a standard, 1D
QIM. The problem of chain partition at finite temperature will be
discussed in detail below.

\subsection{Energies in the subspaces with open Ising chains}

As already mentioned, the general Hamiltonian of the form
(\ref{hamef}) is exactly solvable only in cases when $r_i=r_{i+1}$
or $r_i=-r_{i+1}$ for all $i$. Therefore, one may find exactly the
ground state of spin ladder (\ref{cola}), see below). Otherwise, in
a general case (i.e., in arbitrary subspace) one needs to deal
with a problem of the QIM on an open chain of length $L$ where
$L<N$, described by Hamiltonian (\ref{chain});
\begin{equation}
{\cal H}(\alpha,L)=2J \sum_{i=1}^{L-1}
\tau_{i}^z\tau_{i+1}^z+2J\alpha\sum_{i=1}^L\tau_{i}^x\,.
\label{chagain}
\end{equation}
After applying the JW transformation (\ref{JWT}), Eq.
(\ref{chagain}) takes the form
\begin{eqnarray}
{\cal H}(\alpha,L)&=&2J\sum_{i=1}^L\{(c_{i}^{\dagger}-c_{i})
(c_{i+1}^{\dagger}+c_{i+1})-2\alpha
c_{i}^{\dagger}c_{i}\} \nonumber \\
&+&2JL\alpha\ ,
\end{eqnarray}
with an open boundary condition $c^{\dagger}_{L+1}\equiv 0$. This
condition prevents us from the plane waves expansion, but we can
still use the Bogoliubov transformation. We remark that the broken
chain considered here is sufficient to get a general solution, and
the sum over all subspaces with open (broken) chains is included
in the partition function ${\cal Z}(\alpha)$, see Sec.
\ref{sec:z}.

\begin{figure}[t!]
\includegraphics[width=7cm,angle=0]{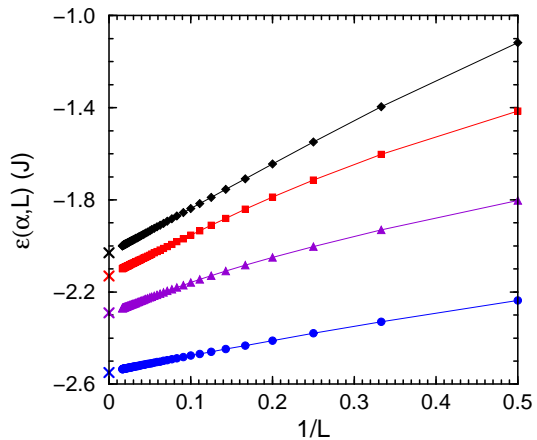}
\caption{(Color online)  Ground state energies per site
$\varepsilon(\alpha,L)$ for the QIM on open chains (\ref{chain})
as functions of inverse chain's length $1/L$ (for $2\leq L\leq
61$) plotted for $\alpha=1$, 3/4, 1/2 and $1/4$, from bottom to
top. Linear fit gives the asymptotic values of energies for
$L\to\infty$, indicated by crosses; these are:
$\varepsilon(\alpha,L\to\infty)=-2.55$, $-2.29$, $-2.13$ and
$-2.03$ for the respective values of $\alpha$. } \label{fig:inf}
\end{figure}

We define new fermion operators $\gamma^{\dagger}_i$ as follows
\begin{equation}
\gamma^{\dagger}_i=\sum_{j=1}^L
\left(\alpha_{ij}c^{\dagger}_j+\beta_{ij}c_j\right)\,,
\end{equation}
for $i=1,2,\dots,L$. Coefficients $\alpha_{ij}$ and $\beta_{ij}$
can be chosen in such a way that the transformation is canonical
and ${\cal H}(\alpha,L)$ takes the diagonal form:
\begin{equation}
{\cal H}(\alpha,L)=\sum_{i=1}^L E_i(\alpha,L)
\left(\gamma^{\dagger}_i\gamma_i-\frac{1}{2}\right)\,.
\label{diagham}
\end{equation}
Both excitations energies $E_i$ and transformation coefficients
$\{\alpha_{ij},\beta_{ij}\}$ can be determined from the condition
\begin{equation}
[{\cal H}(\alpha,L),\gamma^ {\dagger}_i]=E_i\gamma^{\dagger}_i\,.
\label{diagham2}
\end{equation}
This leads to an eigenequation
\begin{equation}
\left(\begin{array}{cc}
A & B\\
-B & -A \end{array}\right) \left( \begin{array}{c}
\vec{\alpha}_{i}\\
\vec{\beta}_{i}\end{array}\right)=E_{i} \left(\begin{array}{c}
\vec{\alpha}_{i}\\
\vec{\beta}_{i}\end{array}\right) \ , \label{matrix}
\end{equation}
where $A$ and $B$ are matrices of size $L\times L$ ($A$ is a
symmetric and $B$ is an antisymmetric matrix), and
$\vec{\alpha}_{i}$, $\vec{\beta}_{i}$ are vectors of length $L$.
The explicit form of $A$ and $B$ for $L=4$ is
\begin{equation}
A=2J\left(\begin{array}{cccc}
-2\alpha & 1 & 0 & 0\\
1 & -2\alpha & 1 & 0\\
0 & 1 & -2\alpha & 1\\
0 & 0 & 1 & -2\alpha\end{array}\right)
\end{equation}
and
\begin{equation}
B=2J\left(\begin{array}{cccc}
0 & 1 & 0 & 0\\
-1 & 0 & 1 & 0\\
0 & -1 & 0 & 1\\
0 & 0 & -1 & 0\end{array}\right) \ ,
\end{equation}
which can be simply generalized to the case of any finite $L$. The
spectrum of ${\cal H}(\alpha,L)$ can be now determined by a
numerical diagonalization of the $2L\times 2L$ matrix from Eq.
(\ref{matrix}). For each $L$ one obtains a set of $2L$ eigenvalues
symmetric around zero. Only the positive ones are the excitation
energies $E_i$ appearing in Eq. (\ref{diagham}). Therefore, the
ground state energy $E_0(\alpha,L)$ is obtained in absence of any
excited states, so the energy per site can be easily expressed as
\begin{equation}
\varepsilon(\alpha,L)=\frac{1}{L}E_0(\alpha,L)
=-\frac{1}{2L}\;\sum_{i=1}^L E_i(\alpha,L) \ . \label{vare}
\end{equation}

Fixing $\alpha$ and increasing $L$ we can trace the dependence of
$\varepsilon(\alpha,L)$ on the system size and make an
extrapolation to an infinite chain $L\to\infty$. Results for
$\varepsilon(\alpha,L)$ (\ref{vare}) as a function of decreasing
$1/L$, obtained for $\alpha=1,3/4,1/2,1/4$ and $L$ changing from
$2$ to $61$, are shown in Fig. \ref{fig:inf}. The energies
decrease with increasing $L$ which suggests that the ground state
corresponds indeed to a closed chain without any defects, as
presented in Sec. \ref{sec:ene}.

\begin{figure}[t!]
\includegraphics[width=7cm,angle=0]{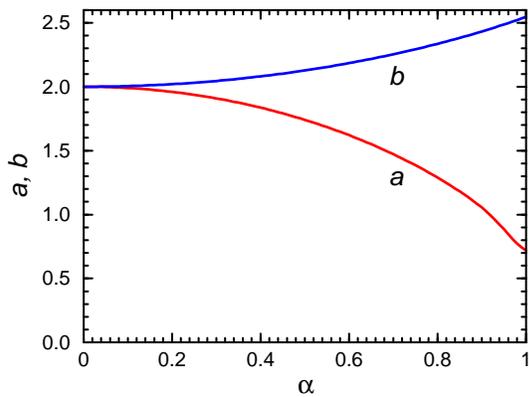}
\caption{(Color online)  Coefficients $a(\alpha)$ (lower line) and
$b(\alpha)$ (upper line) of the linear fit (\ref{appen}) performed
for the points $(1/60,\varepsilon(\alpha,60))$ and
$(1/61,\varepsilon(\alpha,61))$ for different values of $\alpha$.
At $\alpha=0$ one recovers the classical values of the Ising
chain. } \label{fig:ab}
\end{figure}

The dependence of $\varepsilon(\alpha,L)$ on $1/L$ seems to be
almost linear in each case. This is almost exact for $\alpha=1$
and for $\alpha=1/4$, while it holds approximately for
intermediate values of $\alpha$ for in the regime of sufficiently
large $L$. This observation can be used to derive a simple,
approximate formula for the energy $\varepsilon(\alpha,L)$. One
can take the values of $\varepsilon(\alpha,L)$ obtained for two
largest $L$ ($L=60,61$) with fixed $\alpha$ and perform a linear
fit. Hence, we get
\begin{equation}
\varepsilon(\alpha,L)\cong a(\alpha)\,\frac{1}{L}-b(\alpha)
\label{appen}\ ,
\end{equation}
with coefficients $a$ and $b$ depending on $\alpha$. These new
functions can be determined numerically for $\alpha$ changing
between $0$ and $1$ with sufficiently small step. Results obtained
by a numerical analysis are plotted in Fig. \ref{fig:ab}. Both $a$
and $b$ starts from a value $2$ at $\alpha=0$, then $a(\alpha)$
decreases monotonically to about $0.72$ while $b(\alpha)$ slightly
increases to $2.55$ at $\alpha=1$. Eq. (\ref{appen}) is exact for
$\alpha=0$ and any $L$, as well as for $L=60,61$ and any $\alpha$.
Nevertheless, looking at Fig. \ref{fig:inf}, one can expect it to
be a good approximation in case of sufficiently large $L$. From
this formula one can read that for $L\to\infty$ one gets
$E_0(\alpha,L)=-Lb(\alpha)+O(L^0)$ which agrees with the classical
intuition based on extensiveness of the internal energy.

\subsection{Lowest energy excitations}
\label{sec:lowex}

\begin{figure}[t!]
\includegraphics[width=7cm,angle=0]{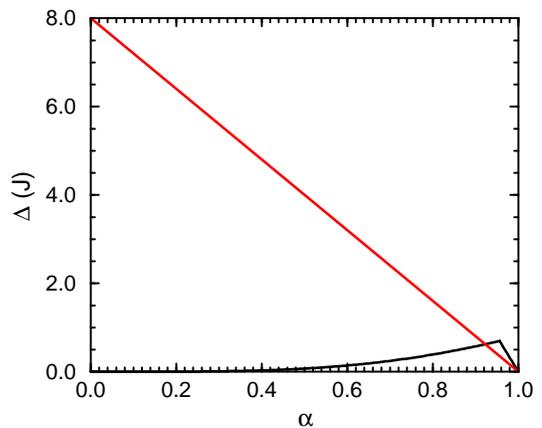}
\caption{(Color online) Excitation energy $\Delta (J)$ as a
function of $\alpha$ for: (i) a ladder with $N=4$ rungs (lower
line) and (ii) an infinite ladder (upper line). In the first case,
as long as the QPT is not approached, the excited state remains in
the ground state subspace with $r_i\equiv -1$ and contains one
Bogoliubov quasiparticle with $k=\pi$. Excitation energy is small,
starts from zero and decreases quickly with growing $N$. The
latter excited state collapses to the ground state for $N=\infty$,
so the first excitation is here different than the one for
infinite $N$. On the contrary, the excited state for $N=\infty$
contains two Bogoliubov quasiparticles with $k=0^{\pm}$. This
leads to the linear gap following $\Delta(\alpha)=8J(1-\alpha)$. }
\label{fig:gap}
\end{figure}

As we pointed out in Sec. \ref{sec:ene}, the lowest excited state
in the case of a finite system, for $\alpha$ far enough from
$\alpha=1$, is simply $\gamma_{\pi}^{\dagger}|0_+\rangle$ and
belongs to the subspace $r_i\equiv -1$. This is a collective
excitation creating a wave of spin--flips in the ground state.
Close to $\alpha=1$ one finds that the lowest excited state is the
ground state from the subspace $r_i\equiv 1$ which means that the
spin order along the rungs changes from AF to FM one along the
$z$--th axis.

The lowest energy excitation changes qualitatively in the
thermodynamic limit $N\to\infty$, where
$\gamma_{\pi}^{\dagger}|0_+\rangle$ and $|0_+\rangle$ states have
the same energy and the dominating excitation is a pair of
Bogoliubov quasiparticles with $k=0^{\pm}$ which corresponds to
flipping one spin at $\alpha=0$. The first excited state remains
in the $r_i\equiv -1$ subspace for all $\alpha$ and the gap
follows linear law $\Delta(\alpha)=8J(1-\alpha)$, see Fig.
\ref{fig:gap}. This shows that in the thermodynamic limit
($N\to\infty$) the low energy spectrum of the ladder is the same
as for ordinary QIM. Note that such behavior is in sharp contrast
with the case of finite ladder of $N=4$ rungs.

\section{Canonical ensemble for the ladder}

\subsection{Partition function}
\label{sec:z}

In order to construct the partition function of spin ladder
(\ref{cola}), we shall analyze its quantum states in different
subspaces. Every invariant subspace introduced in Sec.
\ref{sec:cola} is labelled by a string $r_1r_2\cdots r_N$. Let us
consider an exemplary string of the form
\begin{equation}
1\ 1\ 1\ \bar{1}\ 1\ \bar{1}\ 1\ 1\ \bar{1}\ \bar{1}\ \bar{1}\ 1\
1\ \bar{1}\ 1\ 1\ 1\ 1\ \bar{1}\ \bar{1}\ \bar{1}\ 1\ ,
\label{string}
\end{equation}
where $\bar{1}=-1$, and either $r_i=r_{i+1}$ or $r_i\not=r_{i+1}$.
Each time when $r_i=r_{i+1}$ the chain continues, and when
$r_i\not=r_{i+1}$ we may say that a kink occurs at site $i$ in the
chain. We introduce a periodic boundary condition, so the string
is closed to a loop and $r_N$ stands next to $r_1$. From the point
of view of the reduced Hamiltonian ${\cal H}_{r_{1}r_{2}\cdots
r_{N}}$, given by Eq. (\ref{hamef}), it is useful to split the
string $\{r_i\}$ into chains and kinked areas. A chain is a
maximal sequence of $r_i$'s without any kinks consisting at least
of two sites. Kink areas are the intermediate areas separating
neighboring chains. Using these definitions we can divide our
exemplary string (\ref{string}) as follows
\begin{eqnarray}
1\ 1\ 1)\ \bar{1}\ 1\ \bar{1}\ (1\ 1)\ (\bar{1}\ \bar{1}\
\bar{1})\ (1\ 1)\ \bar{1}\ (1\ 1\ 1\ 1)\ (\bar{1}\ \bar{1}\
\bar{1})\ (1 \ , \nonumber \\
\label{excha}
\end{eqnarray}
where we adopt the convention to denote chains as
$(r_ir_{i+1}\cdots r_{i+p})$, and kink areas as $)r_ir_{i+1}\cdots
r_{i+q}($. For any string of $r_i$'s containing $m$ chains we can
define chain configuration $\{L_i\}$ with $i=1,2,\dots,m$, where
$L_i$'s are the lengths of these chains put in descending order.
In case of our exemplary string its chain configuration is
$\{4,4,3,3,2,2\}$. Variables $\{L_i\}$ must satisfy three
conditions: (i) $L_i\geq 2$ for all $i$, (ii) $\sum_{i=1}^mL_i\leq
N$, and (iii) $\prod_{i=1}^m(-1)^{L_i}\equiv (-1)^m$. The first
two of them are obvious, while the last one is a consequence of
the periodic boundary conditions. Using chain parameters the
effective Hamiltonian ${\cal H}_{r_{1}r_{2}\cdots r_{N}}$ can be
written as a sum of commuting operators
\begin{equation}
{\cal H}_{r_{1}r_{2}\cdots r_{N}}(\alpha) =\sum_{i=1}^m {\cal
H}(\alpha ,L_i) -2J\alpha \sum _{i=1}^K \tau_i^x +2JC_{\vec
r}(\alpha)\ , \label{cham}
\end{equation}
where $K=N-\sum_{i=1}^m L_i$ stands for the total size of kinked
areas. This formula refers to all subspaces excluding those with
$r_{i+1}\equiv r_i$, where we have already obtained exact
solutions. The evaluation of the constant $C_{\vec r}(\alpha)$ can
be completed by considering chain and kink areas in each subspace,
see appendix. Having the diagonal form of ${\cal H}(\alpha,L)$,
given by Eq. (\ref{diagham}), one can now calculate partition
function for the ladder of $2N$ spins. It can be written as
follows
\begin{eqnarray}
{\cal Z}(\alpha)&=&\sum_{\{L_i\}}\sum_{R_{\{L_i\}}}
F_{\alpha}[\{L_i\},R_{\{L_i\}}] e^{-2JC_{\vec r}/T}
Z(\alpha,\{L_i\}) \nonumber \\
&+&Z_0(\alpha)\ , \label{partition}
\end{eqnarray}
where the sum over all $\{\vec r\}$ subspaces is replaced by sums
over all chain configurations $\{L_i\}$ and all $R=\sum_{i=1}^N
r_i$ configurations possible for a given $\{L_i\}$. Factor
$F_{\alpha}[\{L_i\},R_{\{L_i\}}]$ is a number of $\vec r$
subspaces for fixed chain configuration and fixed $R$ when
$\alpha<1$, and for $\alpha=1$ it is a number of $\vec r$
subspaces when only $\{L_i\}$ is fixed. Partition function for any
subspace containing open QIM chains or kinked areas is given by
\begin{eqnarray}
Z(\alpha,\{L_i\})&=&2^N \cosh^K\left[\frac{2J}{T \alpha}\right]
\nonumber \\
&\times&\prod_{i=1}^n \prod_{j=1}^{l_i}\,
\cosh^{N(l_i)}\left[\frac{E_j(\alpha,l_i)}{2T}\right]\ ,
\end{eqnarray}
where $\{l_i\}$ ($i=1,2,\dots,n$) are the different lengths of the
chains appearing in the chain configuration $\{L_i\}$, $N(l_i)$
stands for the number of chains of the length $l_i$, and $T$ is
temperature in units of $k_B=1$. For example, the chain
configuration $\{4,4,3,3,2,2\}$ of Eq. (\ref{excha}) has $n=3$,
$\{l_i\}=\{4,3,2\}$ and $N(l_i)\equiv 2$. The term $Z_0(\alpha)$
is a contribution from subspaces with $r_{i+1}\equiv r_i$. Using
exact solutions (\ref{qimD}), available in these subspaces, one
finds that
\begin{eqnarray}
\!\!Z_0(\alpha)\!&=&\cosh \left[\frac{2J}{T}N(1-\alpha)\right] \nonumber \\
&\times &\!\!\sum_{S=\pm 1}\!\left( \prod_{q=0}^{N-1}\!\cosh
\frac{E_q^S}{T}+ S \prod_{q=0}^{N-1}\!\sinh
\frac{E_q^S}{T}\right), \label{z0}
\end{eqnarray}
where the quasiparticle energies are:
\begin{eqnarray}
E_q^{+}\!&=&2J\left\{1+\alpha^2
+2\alpha\cos\left(\frac{2q+1}{N}\pi\right)\right\}^{1/2},
\\
E_q^{-}\!&=&2J\left\{1+\alpha^2
+2\alpha\cos\left(\frac{2q+2}{N}\pi\right)\right\}^{1/2}.
\end{eqnarray}
Appearance of both sine and cosine hyperbolic functions in $Z_0$
(\ref{z0}) is due to the projection operators ${\cal P}$ introduced
in section \ref{sec:qim}.

\subsection{Combinatorial factor}
\label{sec:kombi}

To obtain numerical values of the partition function one has to
get the explicit form of the combinatorial factor
$F_{\alpha}[\{L_i\},R_{\{L_i\}}]$. This can be done in a simple
way only for $\alpha=1$ when $C_{\vec{r}}(\alpha)=0$, see Eq.
(\ref{cr}). Then we have
\begin{equation}
F_{\alpha=1}[\{L_i\},R_{\{L_i\}}] \equiv F_{1}[\{L_i\}]\ ,
\end{equation}
where $F_{1}[\{L_i\}]$ is the number of different $\vec r$
subspaces that can be obtained from a fixed chain configuration
$\{L_i\}$. Now we can derive a formula for this combinatorial
factor.

The chains can be put into the $r_i$ string in any order and these
of equal length are indistinguishable. Apart from chains, there
are also $r_i$'s belonging to the kinked areas which determine the
actual string configuration. We have $K=N-\sum_{i=1}^m L_i$ of
them, they are indistinguishable and can be distributed among $m$
kinked areas. These degrees of freedom lead to a combinatorial
factor
\begin{equation}
\frac{m!}{N(l_1)!\dots N(l_n)!} \left( \begin{array}{c}
K+m-1\\
K\end{array}\right) ,
\label{kombi}
\end{equation}
where $l_1,l_2,\dots,l_n$ $(n\leq m)$ are the lengths of the
chains without repetitions and $N(l_i)$ is a number of chains of
the length $l_i$. After determining the length of the first chain
$L_1$ and the size of its kink area $A_1$, we still need to fix
the position of $r_1$. We have exactly $L_1+A_1$ possibilities.
Next, we have to sum up over all possible values of $L_1$ (which
are $l_1,l_2,\dots,l_n$), all possible sizes of the kink area
$A_1$ (which are $1,2,\dots,K$) and multiply by a combinatorial
factor (\ref{kombi}) calculated for the remaining part of the
string. The result is
\begin{eqnarray}
F_{1}[\{L_i\}]&=&2\sum_{i=1}^n N(l_i)
\frac{(m-1)!}{N(l_1)!\dots N(l_n)!} \nonumber \\
&\times&\!\sum_{a=0}^{K}\;(l_i+a) \left( \begin{array}{c}
K-a+m-2\\
K-a\end{array}\right), \label{deg1}
\end{eqnarray}
where the factor of $2$ in front comes from the fact that $r_1=\pm
1$. This number tells us how many times a given energy spectrum
repeats itself among all subspaces when $\alpha=1$. The binomial
factor appearing in formula (\ref{deg1}) needs to be generalized
with $\Gamma$ functions when $m=1$.

\section{COMPASS ladder at finite temperature}

\subsection{Correlation functions and chain fragmentation}
\label{sec:mk}

\begin{figure}[t!]
\includegraphics[width=7.2cm,angle=0]{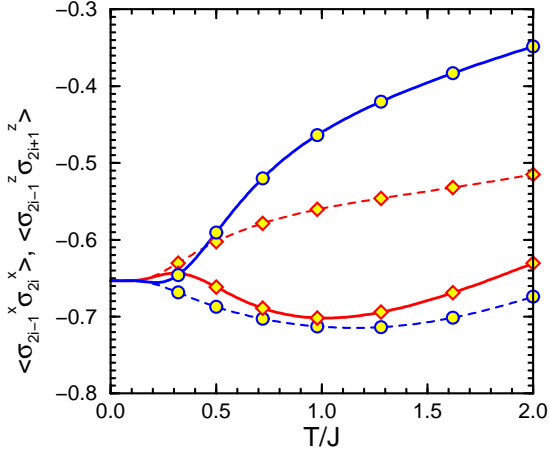}
\caption{(Color online) Nearest neighbor correlation functions,
$\langle\sigma_{2i-1}^x\sigma_{2i}^x\rangle$ on the rungs
(diamonds and red lines) and $\langle \sigma_{2i-1}^z
\sigma_{2i+1}^z\rangle$ along the ladder legs (circles and blue
lines), calculated for the compass ladder ($\alpha=1$) of $2N=8$
spins for increasing temperature $T$, taking into account: ($i$)
all subspaces (solid lines) for increasing temperature $T$, and
($ii$) only the subspace which contains the ground state
$r_i\equiv -1$ (dashed lines).  } \label{fig:cor}
\end{figure}

Nearest-neighbor correlation functions can be easily derived at
finite temperature from the partition function ${\cal
Z}(\alpha,\beta,\gamma)$, if we substitute our initial Hamiltonian
${\cal H}(\alpha)$ given by Eq. (\ref{cola}) by
\begin{eqnarray}
{\cal H}(\alpha,\beta,\gamma)&=&2J \sum_{i=1}^{N}\left\{
 \gamma \sigma^x_{2i-1}\sigma^x_{2i}
+(1-\alpha)\sigma^z_{2i-1}\sigma^z_{2i} \right\} \nonumber \\
&+&J\beta\sum_{i=1}^{N}
\left(\sigma^z_{2i-1}\sigma^z_{2i+1}+\sigma^z_{2i}\sigma^z_{2i+2}
\right)\,.
\end{eqnarray}
Then, after calculating the partition function, we recover spin
correlations by differentiating ${\cal Z}(\alpha,\beta,\gamma)$
with respect to $\beta$ and $\gamma$, and inserting
$\gamma=\alpha$ and $\beta=1$ to the obtained correlations to
derive the final results. Once again, this can be done in a simple
way for small ladders. Correlation functions
$\langle\sigma_{2i-1}^x\sigma_{2i}^x\rangle$ and
$\langle\sigma_{2i-1}^z\sigma_{2i+1}^z\rangle$ for spin ladder
(\ref{cola}) at $\alpha=1$ (quantum compass ladder) are shown in
Fig. \ref{fig:cor} for increasing temperature $T$. Other nearest
neighbor correlations vanish at $\alpha=1$ for trivial reasons.

Fig. \ref{fig:cor} shows the qualitative difference between
correlation functions of spin ladder (\ref{cola}) and those of
periodic QIM chain (\ref{QIM}) of length $N$, that appears in the
ground subspaces $r_i\equiv r_{i+1}$. When all the subspaces are
considered, thermal fluctuations gradually destroy the spin order
along the legs and the
$\langle\sigma_{2i-1}^z\sigma_{2i+1}^z\rangle$ correlations
weaken. On the contrary, the
$\langle\sigma_{2i-1}^x\sigma_{2i}^x\rangle$ correlations on the
rungs are robust in the entire range of physically interesting
temperatures $0<T<2J$, as the ZZ interactions destroying them are
gradually suppressed with increasing $T$ due to the increasing
size of kinked areas.

\begin{figure}[t!]
\includegraphics[width=6.4cm,angle=0]{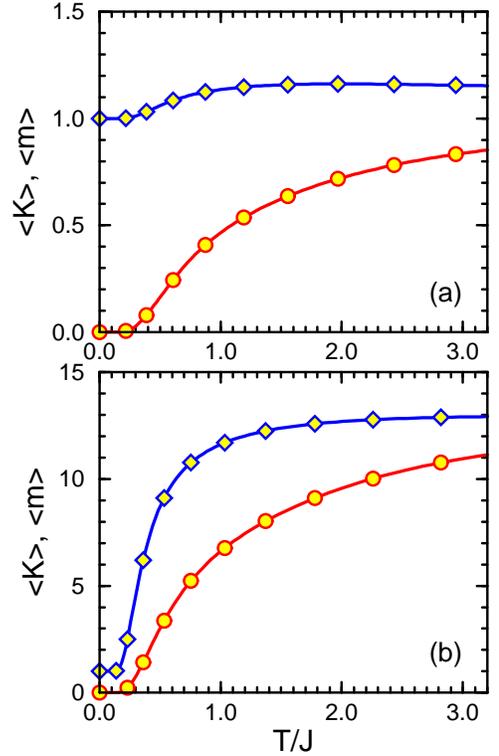}
\caption{(Color online) Average size of the kinked areas $\langle
K\rangle$ (\ref{kav}) (circles and red lines) and the average
number of chains $\langle m\rangle$ (\ref{mav}) (diamonds and blue
lines) for the quantum compass ladder (\ref{cola}) (at $\alpha=1$)
consisting of: (a) $2N=8$, and (b) $2N=104$ spins. The mean size
of kinked areas $\langle K\rangle$ increases monotonically with
increasing temperature $T$ to the asymptotic value $N/4$, see Eq.
(\ref{mkoo}). The average number of chains $\langle m\rangle$
behaves differently, growing quickly to a maximal value at
intermediate $T$ and then decreasing slowly when $T\to\infty$
towards: (a) $1.125$, and (b) $13+12\times 10^{-15}$. }
\label{fig:alldl}
\end{figure}

The above result is qualitatively different from the QIM results
shown by dashed lines in Fig. \ref{fig:cor}, where thermal
fluctuations initially increase intersite correlations of $z$--th
spin components along the ladder legs and reduce the influence of
the transverse field acting on $\tau_i^x$ pseudospins due to spin
interactions $2J\sigma_{2i-1}^x\sigma_{2i}^x$ on the rungs. In the
latter case thermal fluctuation in certain interval of temperature
can enhance local spin ZZ correlations along the ladder legs at
the cost of disorder in the direction of external field. This is
because pseudospin interaction involves $\tau^z_i$ operators, not
$\tau^x_i$ ones. Remarkably, in the full space, see solid lines in
Fig. \ref{fig:cor}, the spin correlations are initially the same
(at low $T$) as those for the QIM, but this changes when
temperature $T\simeq 0.3J$ is reached and the two curves cross ---
then the rung correlations start to dominate. The crossing is
caused by the growth of the kinked areas, as shown in Fig.
\ref{fig:alldl}, which are free of quantum fluctuations and
therefore favor rung correlations of $x$--th spin components.

Another interesting information on excitations in the quantum
compass ladder is the evolution of the average chain configuration
with increasing temperature. As we know from Sec. \ref{sec:z},
every subspace can be characterized by the lengths of chains that
appear in its $r_i$ label. Chain configurations can in turn be
characterized by: ($i$) the number of chains which are separated
by kinks $m$, and ($ii$) the total size of kinked areas $K$.
Thermodynamic averages of both quantities, $\langle m\rangle$ and
$\langle K\rangle$, can be easily determined at $\alpha=1$ even
for a relatively large system using the combinatorial factor
$F_{1}[\{L_i\}]$ (\ref{deg1}) calculated in Sec. \ref{sec:kombi}.
In the limit of $T\to\infty$ one has:
\begin{eqnarray}
\label{mav} \langle m\rangle_{\infty}&=&
\frac{\sum_{\{L_i\}}F_1[\{L_i\}]\left(N-\sum_{j=1}^mL_j\right)}
{\sum_{\{L_i\}}F_1[\{L_i\}]}\,,
\\
\label{kav} \langle K\rangle_{\infty}&=&
\frac{\sum_{\{L_i\}}F_1[\{L_i\}]m[\{L_i\}]}
{\sum_{\{L_i\}}F_1[\{L_i\}]}\,,
\end{eqnarray}
where $m[\{L_i\}]$ is the number of $\{L_i\}$ in the chain
configuration $L_1,L_2,\cdots,L_m$.

In Fig. \ref{fig:alldl} we show the average quantities $\langle
m\rangle$ and $\langle K\rangle$ for ladders of $2N=8$ (left) and
$2N=104$ spins (right). In both cases the average number of chains
$\langle m\rangle$ starts from $1$ and the average size of the
kinked areas $\langle K\rangle$ starts from $0$, corresponding to
a single chain without kinks in the ground state at $T=0$. The
number of chains $\langle m\rangle$ grows to a broad maximum in
the intermediate temperature range and decreases asymptotically to
a finite value. This non--monotonic behavior is due to the fact
that the states with the highest energy, which become accessible
when $T\to\infty$, do not belong to the subspaces with large
number of chains. The mean value of kinks $\langle K\rangle$
follows $\langle m\rangle$ but increases monotonically in the
entire range of $T$, and for finite $T$ one finds that $\langle
K\rangle<\langle m\rangle$. By looking at the current results one
may deduce that in case of $T\to\infty$ and for large $N\gg 1$
both quantities approach
\begin{equation}
\langle m\rangle_{\infty}=\langle K\rangle_{\infty}=\frac{N}{4}\,.
\label{mkoo}
\end{equation}
This is an interesting combinatorial feature of the chain
configurations which is not obvious when we look at the explicit
form of the combinatorial factor $F_{1}[\{L_i\}]$ (\ref{deg1}).
Note that Eq. (\ref{mkoo}) gives an integer due to our choice of
system sizes $2N$ considered here, being multiplicities of 8,
i.e., $N$ is a multiplicity of 4.

\subsection{Spectrum of a large system}
\label{sec:spela}

The combinatorial factor $F_{1}[\{L_i\}]$ given by Eq.
(\ref{deg1}) enables us to calculate the partition function ${\cal
Z}(1)$ (\ref{partition}) for a large system when $\alpha=1$. As a
representative example we consider a ladder consisting of $2N=104$
spins. Even though we can reduce Hamiltonian (\ref{cola}) to a
diagonal form when $2N=104$, as shown in previous paragraphs, it
is still impossible to generate the full energy spectrum for
practical reasons --- simply because the number of eigenstates is
too large. Instead, we can obtain the density of states in case of
$\alpha=1$ using the known form of the partition function
(\ref{partition}) and of the combinatorial factor (\ref{deg1}).
Partition function for imaginary $1/T$ can be written as
\begin{equation}
{\cal Z}(ix)=\sum_{p=0}^{4^N-1}e^{-ix E_p}=\int_{E_0-\varepsilon}^
{-E_0+\varepsilon}dE e^{-ixE} \rho(E)\ ,
\label{zetim}
\end{equation}
where
\begin{equation}
\rho(E)\equiv \sum_{p=0}^{4^N-1}\delta(E-E_p)\ ,
\label{rho}
\end{equation}
and where sum is over all eigenenergies $E_p$ of the ladder.
Parameter $E_0$ is the energy of the ground state. Small and
positive $\varepsilon$ is introduced to formally include $\pm E_0$
into integration interval. Here we used the fact that ladder's
spectrum is symmetric around zero at the compass point $\alpha=1$
(see Fig. \ref{fig:ene}). Function $\rho(E)$ can be easily
recognized as the density of states.

Using $x=2\pi n/w$ in Eq. (\ref{zetim}), with
$w=2(|E_0|+\varepsilon)$ standing for the length of the
integration interval and $n$ being integer, we easily recover the
density of states $\rho(E)$ (\ref{rho}) in a form of the Fourier
cosine expansion
\begin{equation}
\rho(E)=\frac{2}{w}\sum_{n=1}^{\infty} {\cal
Z}\left(2i\pi\frac{n}{w}\right)\cos\left(2\pi\frac{n}{w}
E\right)+\frac{1}{w}{\cal Z}(0)\ , \label{ro}
\end{equation}
with amplitudes given by the partition function ${\cal Z}(ix)$.
%
%
\begin{figure}[t!]
\includegraphics[width=8.2cm,angle=0]{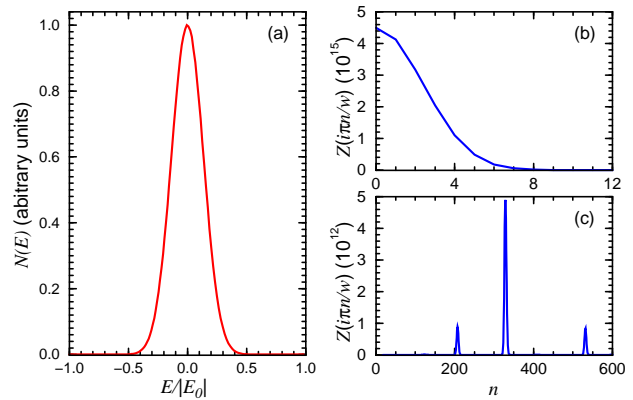}
\caption{(Color online) Relative density of states $N(E)$ (a) as a
function of energy ratio $E/E_0$ and Fourier coefficients ${\cal
Z}\left(2i\pi\frac{n}{w}\right)$ as functions of $n$ for $0\leq
n\leq 12$ (b) and for $13\leq n\leq 600$ (c) calculated for the
ladder of $2N=104$ spins. Relative density of states reminds a
gaussian centered in zero with the width being roughly $0.15$ of
the spectrum width $w$. This follows from the gaussian behavior of
${\cal Z}\left(i2\pi\frac{n}{w}\right)$ coefficients for small $n$
(panel (b)). Plot (c) reveals peaks in ${\cal
Z}\left(i2\pi\frac{n}{w}\right)$ for $n=208,330,533$, three order
of magnitude weaker than for $n=0$, corresponding with periodic
condensations of the energy levels every $\Delta
E=1.28,0.81,0.50J$ (especially every $0.81J$).} \label{fig:dos}
\end{figure}
In practice we cannot execute the sum above up to infinity.
Therefore, it is convenient to define $\rho_c(E)$ which is given
by the same Eq. (\ref{ro}) as $\rho(E)$ but where the sum has a
cutoff for $n=c$. The heights of peaks in $\rho_c(E)$ are expected
to grow in an unlimited way with increasing value of $c$, so it is
convenient to define the normalized density of states $N(E)$ as
\begin{equation}
N(E)=\rho_c(E)/\rho_c(0)\ .
\end{equation}

The results for the compass ladder ($\alpha=1$) of $2N=104$ spins
are shown in Fig. \ref{fig:dos}. These are relative density of
states $N(E)$ for cutoff $c=600$ and Fourier coefficients ${\cal
Z}\left(2i\pi\frac{n}{w}\right)$ for two intervals of $n$. Results
obtained for lower cutoffs show that the overall gaussian shape of
$N(E)$, shown in Fig. \ref{fig:dos}(a), does not change visibly if
only $c>8$. This allows us to conclude that the spectrum of the
compass ladder becomes continuous when the size of the systems
increases which is not the case for the Ising ladder ($\alpha=0$).
Higher values of $n$ are investigated to search for more subtle
effects than gaussian behavior of $N(E)$. These are found by
looking at the amplitudes ${\cal Z}\left(2i\pi\frac{n}{w}\right)$
in high $n$ regime [Fig. \ref{fig:dos}(c)], as the low $n$ regime
[Fig. \ref{fig:dos}(b)] encodes only the gaussian characteristic
of the spectrum. One finds three sharp maxima of the amplitudes
for $n=208,330,533$ out of which the one with $n=330$ is about
five times more intense than the rest, but it is still $10^3$
times weaker than the peak in $n=0$. These values of $n$
correspond with some periodic condensations of the energy levels
with periods $\Delta E=1.28,0.81,0.50J$ respectively which are
visible in $N(E)$ only in vicinity of $E=\pm E_0$.

\section{Heat capacity}
\label{sec:cv}

\subsection{From Ising to compass model}
\label{sec:cv8}

In this Section we analyze heat capacity to identify
characteristic excitation energies in the compass ladder. We begin
with complete results for the ladder consisting of $2N=8$ spins
shown in Fig. \ref{fig:colade}, where all chain configurations can
be written explicitly. Using Eq. (\ref{partition}) for the
partition function, one can next calculate all thermodynamic
functions including average internal energy and the heat capacity.

\begin{figure}[t!]
\includegraphics[width=8.2cm,angle=0]{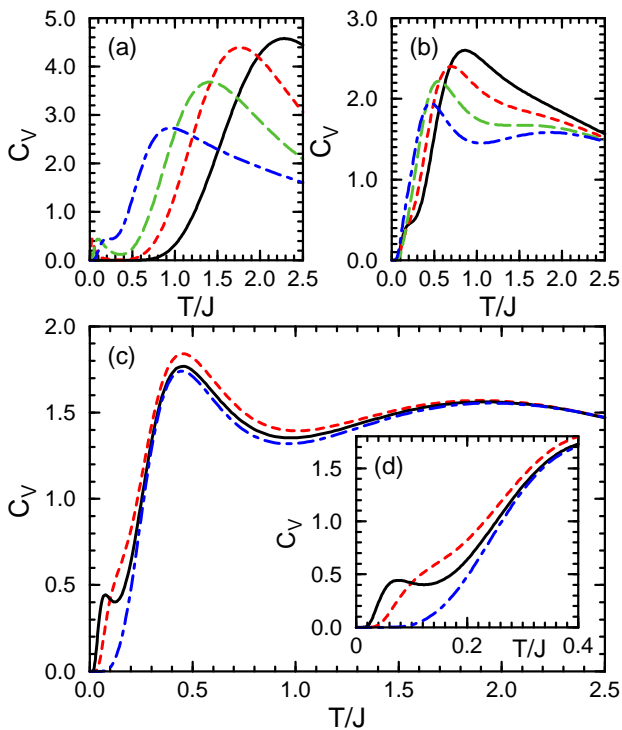}
\caption{(Color online) Heat capacities $C_V$ for spin ladder
(\ref{cola}) of $2N=8$ spins, shown in Fig. \ref{fig:colade}, with
parameter $\alpha$ equal to: (a) $\alpha =0, 0.49, 0.69, 0.85$,
(b) $\alpha =0.87, 0.90, 0.94, 0.97$, and (c) $\alpha =0.982,
0.988, 1$. In panels (a) and (b) lines from right to the left
(solid, dashed, long-dashed, dashed-dotted)
correspond with growing $\alpha$. In panels (c) and (d) the values
of $C_V$ for $\alpha =0.982, 0.988, 1$ are shown by dashed, solid
and dashed-dotted lines, respectively. Panel (d) shows the low
temperature data of panel (c) (for $\alpha>0.98$), with a well
developed small peak at low temperature originating from the
critical excitations between subspaces $r_i\equiv 1$ and
$r_i\equiv -1$ close to $\alpha=1$; it disappears at $\alpha=1$. }
\label{fig:allcv}
\end{figure}

Results for the heat capacity $C_V$ for different values of
$\alpha$ are shown in Fig. \ref{fig:allcv}. These plots cover
three characteristic intervals of $\alpha$ where the behavior of
curves changes qualitatively by appearance or disappearance of
certain maxima. The positions of these maxima correspond to
possible excitation energy scales of the system that change at
increasing $\alpha$ and their intensities reflect the number of
possible excitations in a given energy interval. In case of
$\alpha=0$ [Fig. \ref{fig:allcv}(a)], we see a single maximum at
$\sim 2.2J$ which corresponds to flipping spins in an Ising spin
ladder. Switching on the XX interactions and weakening the ZZ
interactions on the rungs has two effects: ($i$) decreasing energy
and intensities of the high--energy maximum, and ($ii$) appearance
of a low--energy mode in every subspace with QIM chains which
manifests itself as a peak with low intensity at low temperature
$T$, see Fig. \ref{fig:allcv}(a). At $\alpha\simeq 0.85$ this mode
overlaps with modes of higher energies and until $\alpha\simeq
0.94$ there is a single peak again with a shoulder at high values
of $T$, shown in Fig. \ref{fig:allcv}(b). Then the excitation
energies separate again and a broad peak appears for high $T$
accompanied by a distinct maximum at $T\simeq 0.4J$.

In Fig. \ref{fig:allcv} we recognize the characteristic features
for the QIM chains present in most of the subspaces which are
influenced by the excitations mixing different subspaces. If we
had only one subspace with $r_i\equiv -1$, i.e., the one
containing the ground state, then we would have two maxima in
$C_V$ for all $0\leq \alpha\leq 1$ --- one of low intensity in the
regime of low temperature $T$, and another one in high $T$, broad
and intense. The small maximum corresponds with low--energy mode
of QIM that disappears for certain $\alpha>1$. This is not the
case for other subspaces where QIM chains are fragmented and
kinked area are formed. In case of the $1\bar 1\bar 1\bar 1$
subspace the low--energy peak in $C_V$ vanishes at $\alpha\simeq
0.65$ and the high-energy peak persists and moves to higher
temperatures with the increase of $\alpha$. The situation is
similar for the $11 \bar 1\bar 1$ subspace but the peak disappears
at $\alpha\simeq 0.75$ and in the classical subspace $1 \bar 1 1
\bar 1$ we have only one maximum for any $\alpha$. One can deduce
now that the general rule is that the separation of peaks in heat
capacity is reduced primarily by the growth of kinked areas and
secondarily by the fragmentation of chains. This separation of
energy scales is also visible in Fig. \ref{fig:ene} where the
spectra in different subspaces are shown; below certain $\alpha$
in all cases but (d), which is the classical subspace, the energy
gap between the ground state and first excited state is smaller
than other energy gaps appearing in the subspace.

The mixing of different subspaces in the partition function makes
the peaks in $C_V$ overlap which can result in reducing their
number. This happens in Fig. \ref{fig:allcv}(b); for solid
($\alpha=0.87$) and dashed ($\alpha=0.90$) curve we have only one
maximum. For higher or lower $\alpha$ the energy scales remain
separated which is due to fact that: ($i$) soft modes survive in
most of subspaces for low $\alpha$, and ($ii$) for high $\alpha$
the high--energy modes become even tougher and do not overlap with
soft modes still present in subspaces with small kinked areas. The
last phenomenon characteristic for the ladder are excitations
between $r_i\equiv -1$ and $r_i\equiv 1$ subspace in the vicinity
of the QPT. This yields to the appearance of the new energy scale
$\Delta(\alpha)=4NJ(1-\alpha)$ at $\alpha\simeq 0.987$ which
manifests itself as a small peak in heat capacity in low
temperature. This maximum vanishes at $\alpha=1$, as shown in Fig.
\ref{fig:allcv}(d).

\subsection{Generic features at large N}

After understanding the heat capacity in a small system of $N=8$
spin (Sec. \ref{sec:cv8}), we analyze a large system using the
statistical analysis of Sec. VI. Obtaining combinatorial factor
$F_{\alpha}[\{L_i\},R_{\{L_i\}}]$ in case of $\alpha<1$ is
difficult and likely even impossible in a general way without
fixing $N$. Hence we focus on the compass ladder ($\alpha=1$) For
the compass ladder of $2N=104$ spins considered in Sec.
\ref{sec:spela}, one finds $2^{52}$ invariant subspaces. Although
the eigenvalues can be found in each subspace, it is not possible
to sum up over all subspaces for practical reasons and a
statistical analysis is necessary. Therefore, the knowledge of the
combinatorial factor $F_{1}[\{L_i\}]$, see Eq. (\ref{deg1}), is
crucial to calculate partition function ${\cal Z}(1)$
(\ref{partition}). Fortunately, knowing it we only need to
consider different chain configurations which are not very
numerous --- there are only $140854$ of them. This means that on
average each energy spectrum of the effective Hamiltonian repeats
itself almost $32\times 10^9$ times throughout all subspaces.

\begin{figure}[t!]
\includegraphics[width=8cm,angle=0]{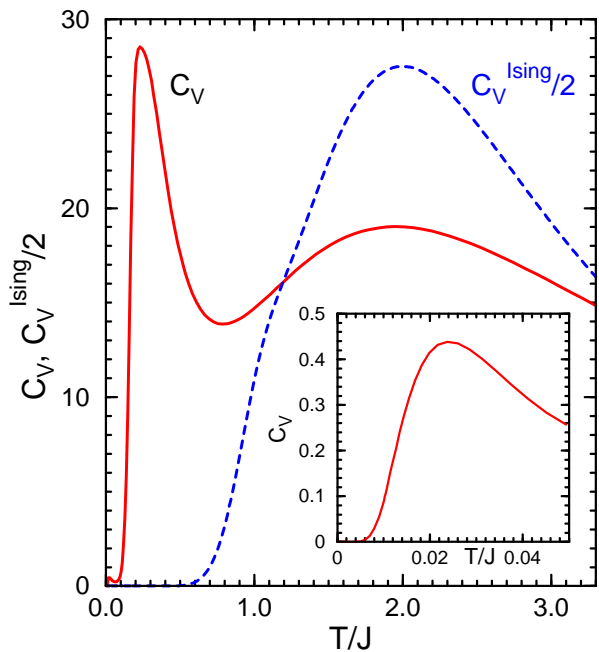}
\caption{(Color online) Heat capacity $C_V$ for the compass ladder
of $2N=104$ spins (spin ladder (\ref{cola}) at $\alpha=1$, solid
line) as a function of temperature $T$ . The main difference with
the case of $2N=8$ spins [see Fig. \ref{fig:allcv}(c)] is a small
maximum appearing at very low $T\simeq 0.02J$, see inset. This
peak originates from the low--energy modes in subspaces $r_i\equiv
r_{i+1}$ which exist in sufficiently long chains described by the
QIM. Dashed line show heat capacity of the Ising ladder
($\alpha=0$) of the same size. } \label{fig:cv52}
\end{figure}

The statistical analysis of the compass ladder consisting of
$2N=104$ spins in terms of: ($i$) mean values of kinked areas
$\langle K\rangle$ (\ref{kav}), and ($ii$) the number of chains
$\langle m\rangle$ (\ref{mav}), was already presented in Fig.
\ref{fig:alldl}(b), while the energy spectrum was discussed in
Sec. \ref{sec:spela}. Here we present the heat capacity $C_V$ for
the compass ladder of this size in Fig. \ref{fig:cv52}. At high
temperature one finds a broad maximum centered at $T\simeq 2J$
which originates from dense excitation spectrum at the compass
point ($\alpha=1$), cf. the spectrum of the compass ladder with
$2N=8$ spins shown in Fig. \ref{fig:ene}. We remark that the broad
maximum of Fig. \ref{fig:cv52} has some similarity to broad maxima
found in the specific heat (heat capacity) of spin
glasses.\cite{Geo01} However, here the broad maximum in the heat
capacity does not originate from disorder but solely indicates
frustration, similar as in some other models with frustrated spin
interactions.\cite{Bec03} We emphasize that the present results
could be obtained only by developing a combinatorial analysis of a
very large number of possible configurations of spin ladder, and
due to the vanishing constant $C_{\vec r}(\alpha=1)=0$ (\ref{cr})
in the energy spectrum for the compass ladder. Unfortunately, the
present problem is rather complex due to the quantum nature of
spin interactions, but in case of the binomial 2D Ising spin glass
an exact algorithm to compute the degeneracies of the excited
states could be developed recently.\cite{Ati08}

The heat capacity $C_V$ of Fig. \ref{fig:cv52} at low temperature
is qualitatively similar to the one obtained for $2N=8$ spins, see
Fig. \ref{fig:allcv}(c), but the steep maximum at low $T$ is here
moved to lower temperature $T\simeq 0.2J$. We also identified an
additional (third) peak in the regime of rather low temperature
$T\simeq 0.02J$ (shown in the inset). This maximum originates from
the QIM (\ref{QIM}) where the energies of the ground state and of
the first excited state approach each other for increasing $N$, if
only $\alpha\leq 1$. Thus, this lowest peak in the heat capacity
obtained for the compass ladder of $2N=104$ spins has to be
considered as a finite size effect --- for increasing system size
it is shifted to to still lower temperature $T$, and would
disappear in the thermodynamic limit $N\to\infty$, in agreement
with the qualitative change of low energy spectrum of the QIM.

\section{Summary and conclusions}
\label{sec:summa}

We have investigated an intriguing case of increasing frustration
in a spin ladder (\ref{cola}) which interpolates between the
(classical) Ising ladder and the frustrated compass ladder when
the parameter $\alpha$ increases from $\alpha=0$ to $\alpha=1$.
The ground state of the ladder was solved exactly in the entire
parameter range by mapping to the QIM, and we verified that
frustrated interactions on a spin ladder generate a QPT at
$\alpha=1$, when conflicting interactions ZZ along the ladder legs
compete with 2XX ones along the rungs. At this point the spin
correlations on the rungs
$\langle\sigma^z_{2i-1}\sigma^z_{2i}\rangle=-1$ collapses to zero
and the ground state becomes disordered. We have shown that the
ground state of a finite ladder has then degeneracy 2, while the
analysis of the energy spectra for increasing size suggests that
the degeneracy increases to 4 in the thermodynamic limit. We note
that this result agrees with degeneracy $2\times 2^L$ found for
the 2D compass model,\cite{Mil05} where $L$ is a linear dimension
(the number of bonds along one lattice direction) of an $L\times
L$ cluster in the 2D system. In our case of a $2\times N$ ladder,
$L=1$ for ladder rungs, so indeed the degeneracy is $2\times 2=4$.

The present method of solving the energy spectrum in different
subspaces separately elucidates the origin of the QPT found in the
present spin ladder (\ref{cola}) at the point $\alpha=1$,
corresponding to the frustrated interactions in the compass
ladder. We argue that this approach could help to find exact
solutions in a class of quasi-1D models with frustrated spin
interactions, but in some cases only the ground state and not the
full spectrum can be rigorously determined. For instance, this
applies to a spin ladder with frustrated spin interactions between
different triplet components on the rungs,\cite{Brz08} where a
different type of a QPT was found recently.

By performing a statistical analysis of different possible
configurations of spin ladder (\ref{cola}) with periodic boundary
conditions we derived a partition function ${\cal Z}(\alpha)$ for
a mesoscopic system of 104 spins. The calculation involves the
classification of ladder subspaces into classes of chain
configurations $\{L_i\}$ equivalent by symmetry operations and the
determination of the combinatorial factor
$F_{\alpha}[\{L_i\},R_{\{L_i\}}]$. We have shown that this factor
can be easily determined at the compass point ($\alpha=1$), so the
heat capacity of such a mesoscopic compass ladder could be found.

Summarizing, we demonstrated that spin ladder studied in this
paper exhibits a QPT from a classical ordered to a quantum
disordered ground state which occurs due to the level crossing,
and is therefore of first order. It leads to a discontinuous
change of spin correlations on the rungs when the interactions
along the ladder legs and on the rungs become frustrated.
Fortunately, the subspaces which are relevant for the QPT in the
compass ladder considered here can be analyzed rigorously, which
gives both the energy spectra and spin correlation functions by
mapping the ladder on the quantum Ising model. The partition
function derived in this work made it possible to identify the
characteristic scales of excitation energies by evaluating the
heat capacity for a mesoscopic system.

{\it Note added in proof.\/} After this paper was accepted, we
learned about a powerful algebraic method to analyze exactly
solvable spin Hamiltonians.\cite{Nus09} The present quantum
compass laddeed could be also analyzed using this approach.

\acknowledgments

We acknowledge support by the Foundation for Polish Science (FNP)
and by the Polish Ministry of Science and Higher Education under
Project No.~N202 068 32/1481.

\appendix*
\section{Evaluation of the energy origin $C_{\vec r}(\alpha)$
in invariant subspaces}
\label{sec:appendix}

We need to express $\sum_{i=1}^N r_i$, which appears in $C_{\vec
r}(\alpha)$, see Eq. (\ref{cr}), in terms of chain configurations
$\{L_i\}$. This task may be accomplished by the following
construction. Let us imagine certain string of $r_i$'s written in
terms of chains $\{L_i\}$ and kink areas $\{A_i\}$:
\begin{equation}
A_1(L_1)A_2(L_2)A_3(L_3)\cdots A_k(L_k)\ .
\nonumber
\end{equation}
First, we want to calculate the sum of $r_i$'s included in chains.
We choose any $r_i$ from the chain $L_1$ and fix its sign as
$r_{\rm in}$. Now this chain gives $r_{\rm in}L_1$ contribution to
the total sum of $r_i$'s. To get to the second chain we have to
pass through the first kink area $A_1$. If the number of kinks in
$A_1$ is even, then the next chain will give the contribution
$r_{\rm in}L_2$, and if not, then it will give the opposite
number. Therefore, after passing through the whole system we will
get the term
\begin{equation}
r_{\rm in}(L_1+p_2L_2+p_2p_3L_3+\dots +p_2p_3\dots p_kL_k)\ ,
\label{rin}
\end{equation}
where $p_i=(-1)^{K_i}$, and $K_i$ is a number of kinks in kink
area $i$. It is clear that the parameters $\{p_i\}$ satisfy
$\prod_{i=1}^k p_i\equiv 1$. Now we need to calculate the sum of
$r_i$'s placed in kink areas. The sign of the first chain is
already chosen as $r_{\rm in}$ so we pass to $A_2$. For even
number of kinks in $A_2$ the contribution is zero. If the number
is odd, then we get the sum equal $-r_{\rm in}$. Passing to the
next kink area we follow the same rules but we have to change
$r_{\rm in}$ into $p_2r_{\rm in}$. The total contribution from the
kink areas is then equal to
\begin{equation}
-p_1r_{\rm in}\left\{\frac{1+p_1}{2}+ \sum_{i=2}^k p_1p_2\dots
p_{i-1}\frac{1+p_i}{2}\right\} \ . \label{p1rin}
\end{equation}
Using the results given in Eqs. (\ref{rin}) and (\ref{p1rin}) we
obtain finally
\begin{equation}
\sum _{i=1}^N r_i=r_{\rm in}\left\{ L_1-1+ \sum _{i=2}^k
p_2p_3\dots p_i(L_i-1)\right\}\ . \label{ppr}
\end{equation}
Thanks to this result, we can write the energy given by Eq.
(\ref{cham}) in terms of new variables $\{L_i,p_i\}$ instead of
$\{r_i\}$ which are definitely more natural for the present
problem.

\end{document}